\documentclass[preprint2]{aastex631}

\usepackage{comment}
\usepackage{amsmath}

\begin{document}

\title{Probing the Galactic Diffuse Continuum Emission with COSI}

\author[0000-0002-6774-3111]{Christopher M. Karwin}
\email{christopher.m.karwin@nasa.gov}
\affiliation{NASA Postdoctoral Program Fellow}
\affiliation{NASA Goddard Space Flight Center, Greenbelt, MD, 20771, USA}

\author[0000-0002-0552-3535]{Thomas Siegert}
\affiliation{Julius-Maximilians-Universität Würzburg, Fakultät für Physik und Astronomie, Institut für Theoretische Physik und Astrophysik, Lehrstuhl für Astronomie, Emil-Fischer-Str. 31, D-97074 Würzburg, Germany}

\author[0000-0002-0128-7424]{Jacqueline Beechert}
\affiliation{Space Sciences Laboratory, UC Berkeley, 7 Gauss Way, Berkeley, CA 94720, USA}

\author[0000-0001-5506-9855]{John A. Tomsick}
\affiliation{Space Sciences Laboratory, UC Berkeley, 7 Gauss Way, Berkeley, CA 94720, USA}

\author[0000-0002-2621-4440]{Troy A. Porter}
\affiliation{W.W. Hansen Experimental Physics Laboratory and Kavli Institute for Particle Astrophysics and Cosmology, Stanford University, Stanford, CA 94305, USA}

\author[0000-0002-6548-5622]{Michela Negro}
\affiliation{Department of Physics and Astronomy, Louisiana State University, Baton Rouge, LA 70803 USA}

\author[0000-0001-6677-914X]{Carolyn Kierans}
\affiliation{NASA Goddard Space Flight Center, Greenbelt, MD, 20771, USA}

\author[0000-0002-6584-1703]{Marco Ajello}
\affiliation{Department of Physics and Astronomy, Clemson University, Clemson, SC 29634, USA}

\author[0000-0002-2471-8696]{Israel Martinez Castellanos}
\affiliation{NASA Goddard Space Flight Center, Greenbelt, MD, 20771, USA}
\affiliation{Department of Astronomy, University of Maryland, College Park, Maryland 20742, USA}

\author[0000-0001-6874-2594]{Albert Shih}
\affiliation{NASA Goddard Space Flight Center, Greenbelt, MD, 20771, USA}

\author[0000-0001-9067-3150]{Andreas Zoglauer}
\affiliation{Space Sciences Laboratory, UC Berkeley, 7 Gauss Way, Berkeley, CA 94720, USA}

\author[0000-0001-9567-4224]{Steven E. Boggs}
\affiliation{Department of Astronomy \& Astrophysics, UC San Diego, 9500 Gilman Drive, La Jolla CA 92093, USA}

\collaboration{12}{(for the COSI Collaboration)}



\begin{abstract}

In 2016 the Compton Spectrometer and Imager (COSI) had a successful 46-day flight onboard NASA's Super Pressure Balloon platform. In this work we report measurements of the Galactic diffuse continuum emission (GDCE) observed towards the inner Galaxy during the flight, which in the COSI energy band ($0.2 - 5$ MeV) is primarily generated from inverse Compton radiation. Within uncertainties we find overall good agreement with previous measurements from INTEGRAL/SPI and COMPTEL. Based on these initial findings, we discuss the potential for further probing the GDCE with the 2016 COSI balloon data, as well as prospects for the upcoming satellite mission.   

\end{abstract}



\section{Introduction} \label{sec:intro}
The Galactic plane is a bright source of diffuse $\gamma$-ray emission, consisting of both continuum and line components. For energies $\lesssim$70~MeV the diffuse emission is mainly due to the interactions of cosmic ray (CR) electrons and positrons with the interstellar gas and radiation fields. The major contribution is by Compton scattering of the CR leptons with the interstellar radiation field (ISRF), which is the result of stellar emission and re-processing of the starlight by dust in the diffuse interstellar medium (ISM), and the CMB~\citep{2005ICRC....4...77P,2008ApJ...682..400P}. There are also sub-dominant contributions from Bremsstrahlung interactions with the gas, and multi-channel positron annihilation~\citep{1971NASSP.249.....S,2022PhRvD.106b3030B}. Observations of the GDCE in the MeV band thus provide a unique probe into the nature of both primary and secondary CR electrons. More specifically, the observations provide crucial information about the electron spectrum throughout the Galaxy, which in turn has implications for the electron source density, injection, and propagation. One key aspect of this is identifying the different source classes of CR electrons, which has remained a longstanding open issue in astrophysics~\citep{1994A&A...292...82S}. Probing the GDCE in the MeV range provides especially useful information for the sub-GeV CR electron spectrum. Direct measurements unaffected by the solar modulation are only available for energies $\lesssim$50 MeV, taken by the \textit{Voyager}~1 spacecraft that has been observing in the local ISM since 2012~\citep{2016ApJ...831...18C}. The MeV $\gamma$-ray data enable us to fill this gap in energy sampling of the CR spectrum, where crucial information about primary CR lepton sources may be accessed. In contrast to the MeV band, the GDCE observed at high energies by the \textit{Fermi} Large Area Telescope (\textit{Fermi}-LAT) is dominated by hadronic processes producing mesons that decay to $\gamma$ rays, which makes disentangling the contribution to the $\gtrsim$100~MeV diffuse emission by CR leptons more challenging.

At MeV energies the Galactic diffuse emission also has significant contributions from many prominent line sources. Among these sources is positron annihilation, as mentioned above~\citep[also see][]{1973ApJ...184..103J,1978ApJ...225L..11L,Kierans:2019aqz,Siegert:2020oxw}; the $^{26}$Al nuclear decay line at 1.809 MeV~\citep{mahoney1984heao,Beechert:2022myg}; and the $^{60}$Fe nuclear decay lines at 1.173 and 1.333 MeV~\citep{2007A&A...469.1005W,2020ApJ...889..169W}. Given the diffuse nature of these line sources, there can be a strong correlation when analyzing the GDCE and line emission, even though the different components are not necessarily expected to share the same morphology. 

Observations of the GDCE have long been hindered by a high level of source confusion in the Galactic plane, due to the large number of point sources that reside there~\citep{2008ApJ...679.1315B,2008ApJ...682..400P,2011ApJ...739...29B}. For $\gamma$-ray detectors in general, the capacity of any given instrument to disentangle point source emission from truly diffuse emission is ultimately limited by the spatial resolution and sensitivity of the instrument. This inevitably leads to some level of contribution from unresolved point sources for any determination of the diffuse emission, and likewise, it can result in significant systematic uncertainties~\citep{2007A&A...463..957K,2008ApJ...679.1315B}.  

Further limiting the capacity to study the GDCE in the MeV band is the fact that the energy range between $\sim0.1-30$ MeV is one of the least explored regions of the electromagnetic spectrum (i.e.~the so-called MeV gap), where current instruments have a continuum sensitivity that is significantly worse than the adjacent X-ray and high-energy $\gamma$-ray bands. There are two primary reasons for this gap in sensitivity. First, the dominant interaction of MeV $\gamma$ rays in matter is Compton scattering, and such events require sophisticated analysis techniques to accurately reconstruct~\citep{2000A&AS..145..311B,2006PhDT.........3Z}. Second, the instrumental background is very high due to the activation of irradiated materials in space, which limits the sensitivity~\citep{2019MmSAI..90..226C}.   
 
Despite these challenges, the GDCE in the MeV band has been probed by numerous satellite experiments. This includes measurements between $1 - 30$ MeV by the Imaging Compton Telescope (COMPTEL)~\citep{1994A&A...292...82S,1996A&AS..120C.381S,2011crpa.conf..473S,Strong:2019yfj}, and between $0.05 - 10$ MeV by the Oriented Scintillation Spectrometer Experiment (OSSE)~\citep{1995ICRC....2..219S,1997ApJ...483L..95S,1999ApJ...515..215K}, both instruments onboard the Compton Gamma Ray Observatory (CGRO)~\citep{1993ApJS...86..657S}. Measurements have also been made between $0.02-2.4$ MeV by the Spectrometer onboard the International Gamma-Ray Astrophysics Laboratory (INTEGRAL/SPI)~\citep{2008ApJ...679.1315B,2008ApJ...682..400P,2011ApJ...739...29B}, with updated measurements recently reported in~\citet{2022A&A...660A.130S} and~\citet{2022PhRvD.106b3030B}, which achieved an improved signal-to-noise ratio and an extension of the spectrum up to 8 MeV. Additionally, observations were made between $0.3 - 8.5$ MeV by the Solar Maximum Mission~\citep[SMM,][]{1990ApJ...362..135H}. A number of balloon-borne measurements have been made as well, including between $0.02 - 10$ MeV by the Gamma-Ray Imaging Spectrometer (GRIS)~\citep{1985ICRC....3..307T,1993ApJ...407..597G}, and between $30 - 800$ keV by the High Resolution Gamma-Ray and Hard X-Ray Spectrometer (HIREGS)~\citep{2000ApJ...544..320B}. The current state of these measurements indicates an excess signal above model predictions towards the Galactic center by a factor of $\sim2-3$~\citep{2008ApJ...682..400P,2011ApJ...739...29B,2011crpa.conf..473S,2018MNRAS.475.2724O,2022A&A...660A.130S,2022arXiv221205713T}. This has been attributed to the possibility of an enhanced ISRF; modification of the CR source density, injection, and/or diffusion; or the presence of a sub-threshold point source population. 

Overall, there are still numerous open questions pertaining to the fundamental nature of CR electrons in the Galaxy and their associated diffuse emission, and this warrants further investigation from new and independent probes. One such probe now on the horizon is the Compton Spectrometer and Imager (COSI), a Small Explorer satellite mission recently selected by NASA and scheduled to launch in 2027~\citep{COSIofficial,BeechertCOSICalib,2023arXiv230812362T}. In 2016 COSI was flown as a balloon-borne experiment onboard NASA's Super Pressure Balloon platform, and had a successful 46-day flight, launching from Wanaka, New Zealand, and landing in Peru~\citep{Kierans:2016qik}. COSI primarily detects photons between $0.2 - 5$ MeV via Compton scattering, and it functions as a spectrometer, wide-field imager, and polarimeter. COSI has already detected emission from the Crab pulsar~\citep{Zoglauer:2021coa}, the Galactic 0.511 MeV line~\citep{Kierans:2019aqz,Siegert:2020oxw}, the Galactic $^{26}$Al nuclear decay line at 1.809 MeV~\citep{Beechert:2022myg}, and a $\gamma$-ray burst~\citep{2017ApJ...848..119L}, as well as likely detections of Cyg X-1, Cen A, and a number of other sources~\citep{Kierans:2016qik}. More details about COSI's main science goals, the instrument, and the 2016 balloon flight can be found in~\citet{Kierans:2016qik},~\citet{COSIofficial},~\citet{BeechertCOSICalib},~\citet{2023arXiv230812362T} and references therein. 

In preparation and anticipation for the COSI satellite mission, the COSI balloon data provide an ample opportunity to make a unique probe of the GDCE in the MeV band. Here we report measurements of the total diffuse continuum flux observed towards the inner Galaxy. A follow-up analysis to quantify the full range of systematic uncertainties associated with the background subtraction is currently underway and will be reported in a forthcoming work. Our analysis proceeds as follows. In Section~\ref{sec:analysis} we describe the analysis, including the data extraction, GALPROP models, MEGAlib simulations, and flux calculation. Results are presented in Section~\ref{sec:results}. In Section~\ref{sec:conclusion} we give our summary, discussion, and conclusions. Additional details about our atmospheric corrections are provided in Appendix~\ref{sec:appendix}. 

\section{Analysis} \label{sec:analysis}
\subsection{2016 COSI Balloon Data} Data for the 2016 COSI balloon flight is taken from~\citet{Beechert:2022myg}, which analyzed the Galactic $^{26}$Al nuclear decay line at 1.809 MeV. Although focused on the detection of $^{26}$Al, a residual continuum signal towards the inner Galaxy was also found in that work. In this section we summarize the key points of the data extraction, but the reader is referred to~\citet{Beechert:2022myg} for full details. 

The signal region is centered on the Galactic center with corresponding Galactic coordinates $|l| \leq 30^\circ$ and $|b| \leq 10^\circ$. This defines cuts in the zenith pointing of the instrument. Additionally, the Compton scattering angle effectively broadens the observation region, and a maximum Compton scattering angle of $\phi_{\mathrm{max}}=35^\circ$ was used. The background region covers the remaining extent of the sky such that the extensions from the maximum Compton scattering angle beyond the borders of the signal and background regions do not overlap. Observations of the signal region are limited to balloon altitudes of at least 33 km, in order to mitigate worsening atmospheric background and attenuation with decreasing balloon altitude. The only times disregarded in the background region are those before the balloon reached float altitude and those with extraneous high rates in the CsI anti-coincidence detectors flown on the balloon instrument. To account for changes in the geomagnetic rigidity during the flight, the data was binned into four geomagnetic latitude bins, each weighted by the respective exposure time. The analysis is also split into two parts due to a high-voltage detector failure that occurred during the flight: data and simulations prior to 2016 June 6 are processed with a 10-detector mass model and afterwards with a 9-detector mass model.  

\subsection{GALPROP Models} 
The GDCE is modeled using the v57 release of the GALPROP CR propagation and interstellar emissions framework~\citep{2022ApJS..262...30P}. GALPROP self-consistently calculates spectra and abundances of Galactic CR species and associated diffuse emissions ($\gamma$-rays, X-rays, and radio) in 2D and 3D. We employ the steady-state emissions examples supplied with the v57 GALPROP release to estimate a likely variation from models that reproduce the latest CR data. There are six models in total, categorized according to the CR source and ISRF model used for the prediction. There are 3 CR source models (SA0, SA50, SA100) and two ISRF models (R12, F98). The CR source density models are based on the distribution of injected CR power, with SA0 describing an axisymmetric disk (following the radial distribution of pulsars), SA50 describing a 50/50\% split of the injected CR luminosity between disk-like and spiral arms, and SA100 describing pure spiral arms. All models have the same exponential scale height of 200 pc. The two ISRF models employ different spatial densities for both the stars and the dust but produce intensities very similar to those of the data for near- to far-infrared wavelengths at the location of the solar system~\citep[see][and references therein]{porter2017high}. For the neutral gas distributions (atomic and molecular), a 3D model from~\cite{Johannesson:2018bit} is employed. Note that these GALPROP models include the total emission, which is dominated by IC, but also has a small contribution from Bremsstrahlung towards the upper energy bound. As our representative case for the response calculation we use the SA100-F98 model, as described in Section~\ref{sec:aeff}. 

In addition to our representative model, we also use two additional GALPROP models from~\citet{2022A&A...660A.130S}, in which updated measurements of the GDCE towards the inner Galaxy were recently made with INTEGRAL/SPI. Specifically, we use the baseline model from that work~\citep{2019ApJ...878...59B}, as well as the model that best-matched the data. These models only include IC emission. Note that the model from~\citet{2019ApJ...878...59B} is based on a combined study of \textit{Voyager} 1 and PAMELA CR data. On the other hand, the parameters for the best-matched model are not physically motivated, but rather they are tuned to enable a good agreement with the INTEGRAL/SPI observations.

\subsection{MEGAlib Simulations} 
\label{sec:mega}
COSI's response to the GDCE is simulated using the Medium-Energy Gamma-ray Astronomy library (MEGAlib) software package\footnote{\url{https://megalibtoolkit.com/home.html}}, a standard tool in MeV astronomy~\citep{2006NewAR..50..629Z}. MEGAlib is based on GEANT4~\citep{geant4_2003} and is able to simulate the emission from a source, the passage of particles through the spacecraft, and their energy deposits in the detector. It also performs event reconstruction, imaging analysis, general high-level analysis (i.e.~spectra, light curves, etc.), and allows for estimates of the background rates from different expected components. 

For running the simulations we employ the COSI simulation pipeline\footnote{\url{https://github.com/cositools/cosi-data-challenges}}, which is available in COSItools (currently being developed for the COSI mission). The simulation pipeline provides a python wrapper for MEGAlib, as well as numerous auxiliary functions, and a growing library of available astrophysical sources. Simulations are run for the full 46-day COSI balloon flight, and they closely mimic the real time-dependence of the instrument's pointing on the sky. The transmission probability is included, which accounts for photons that scatter and don't reach the detector. For the transmission probability we assume an altitude of 33.5 km, which is representative of the average value during the flight. The simulations are split into two time periods (before and after 2016 June 6), corresponding to the same 9-detector and 10-detector mass models used in the data extraction, as discussed above. Correspondingly, when extracting the simulated photons we use the same exact event selections as was used for the real data extraction with \textit{mimrec} (MEGAlib's high-level analysis tool). With this setup we simulate 10 realizations of the GDCE for each of our GALPROP models.

\begin{figure}[t]
\centering
\includegraphics[width=1\columnwidth]{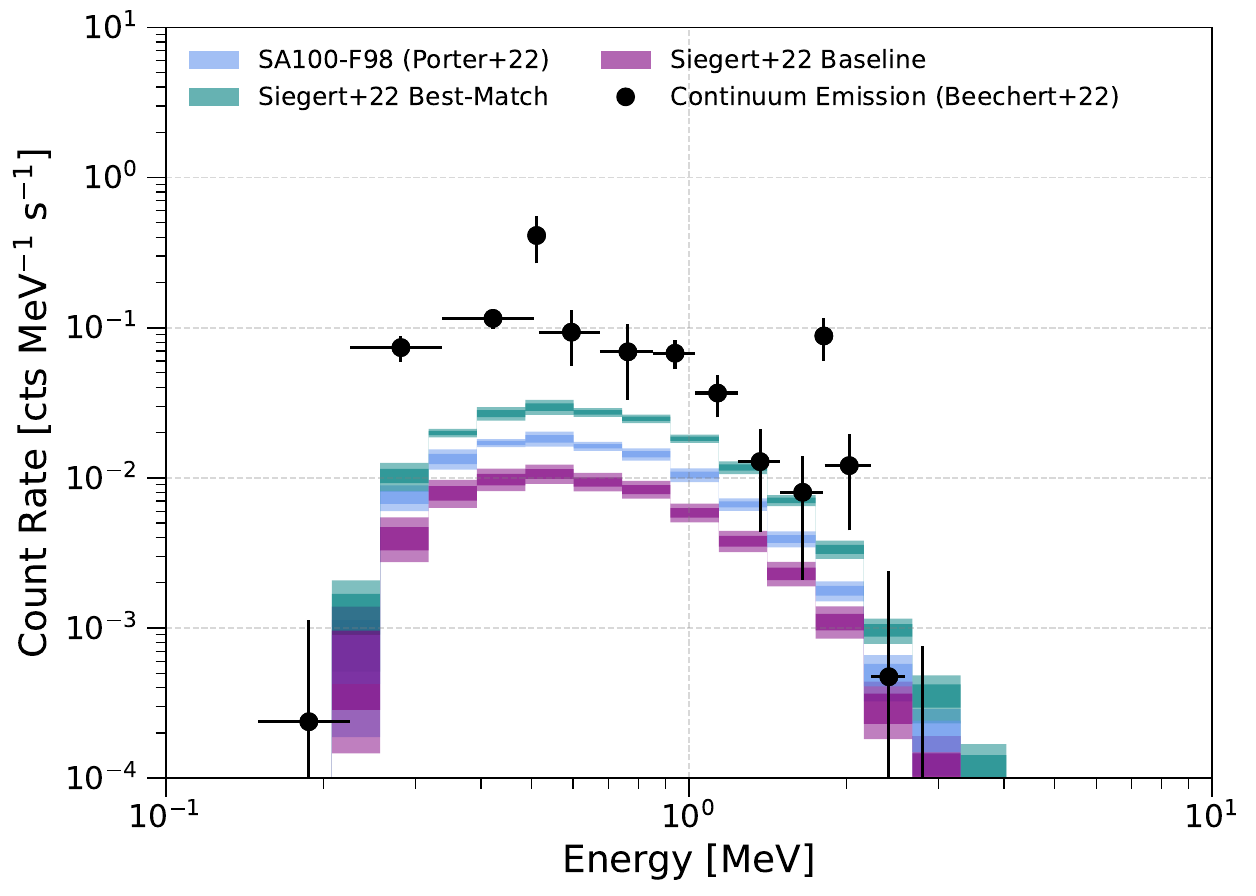}
\caption{The black data points show the continuum emission extracted from~\citet{Beechert:2022myg}. The error bars are statistical only. The colored bands show the expected count rates for three different GALPROP models of the Galactic diffuse emission. Note that the GALPROP models do not include the line emissions. The predictions are based on MEGAlib simulations, which closely replicate the 2016 COSI observations (i.e.~detector and pointing on the sky). The dark and light regions show the $1\sigma$ and $2\sigma$ statistical uncertainties, respectively.}
\label{fig:sims}
\end{figure}

The expected count rates from the MEGAlib simulations are shown in Figure~\ref{fig:sims}. The residual emission resulting from the subtraction of the signal and weighted background region from~\citet{Beechert:2022myg} is shown with black data points, and the corresponding data is provided in Appendix~\ref{sec:data_tables}. As can be seen, the observed emission is systematically higher than the expectations from the simulations. However, in order to make a direct comparison to previous measurements, the count rate must be converted to flux. Additionally, although the simulations already include the transmission probability, this does not account for source photons that get scattered into the detector. These two factors are discussed further in the next section.

\subsection{Effective Area and Flux Conversion}
\label{sec:aeff}
\begin{figure}[t]
\centering
\includegraphics[width=1\columnwidth]{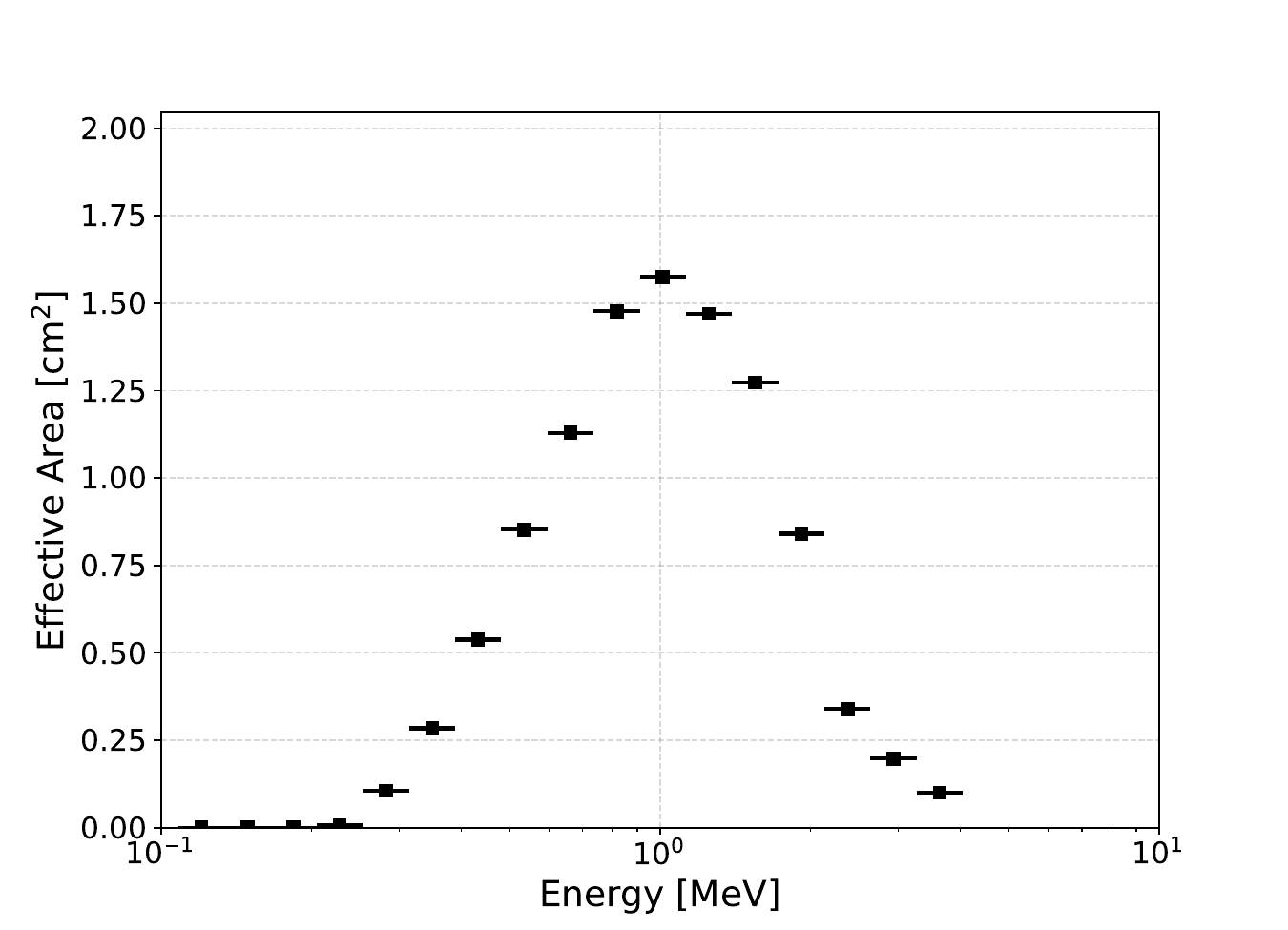}
\caption{The effective area for the Galactic diffuse emission (SA100-F98) calculated from MEGAlib simulations. This represents the average from the total exposure time and is used to convert the observed count rate to flux.}
\label{fig:aeff}
\end{figure}

We convert the observed count rate from~\citet{Beechert:2022myg} to flux by scaling it by the effective area, similar to approaches that have been employed in the past~\citep{2012ApJ...751..108A,Caputo:2022xpx}. The effective area ($A_{\mathrm{eff}}$) for a given energy bin is calculated from the MEGAlib simulations using the formulation
\begin{equation}
    A_{\mathrm{eff}} = \frac{N_{\mathrm{det}}}{N_{\mathrm{sim}}{}} \times A_{\mathrm{ref}},
\end{equation}
where $N_{\mathrm{det}}$ and $N_{\mathrm{sim}}$ are the detected and simulated counts, respectively, and $A_{\mathrm{ref}}$ is a reference area in the simulations equal to 11310 $\mathrm{cm^2}$. The number of simulated counts in a given energy bin is calculated as
\begin{equation}
    N_{\mathrm{sim}} = \bar{F}_{\mathrm{reg}} \times  \Delta t \times A_{\mathrm{ref}} \times \Delta \Omega \times \Delta E, 
\end{equation}
where $\bar{F}_{\mathrm{reg}}$ is the average model intensity over the region of the sky covered by the observations (in units of $\mathrm{ph \ cm^{-2} \ s^{-1} \ sr^{-1} \ MeV^{-1}}$), $\Delta t$ is the total exposure time, $\Delta \Omega$ is the corresponding solid angle of the sky, and $\Delta E$ is the width of the given energy bin. Our signal region is spanned by $|l| \leq 30^\circ$ and $|b| \leq 10^\circ$, plus a Compton broadening of $\phi_{\mathrm{max}}=35^\circ$ in all directions, which results in a sky region of $\Delta \Omega = 130^\circ \times 90^\circ = 3.21 \ \mathrm{sr}$ (obtained from integration). The resulting energy-dependent effective area is shown in Figure~\ref{fig:aeff}. This specific calculation is for the SA100-F98 model, but all models give comparable results. Note that the simulations replicate the time-variation of the detector's pointing during the observations, and thus the effective area calculated here represents the average over all pointings. 

\begin{figure*}[t]
\centering
\includegraphics[width=0.64\textwidth]{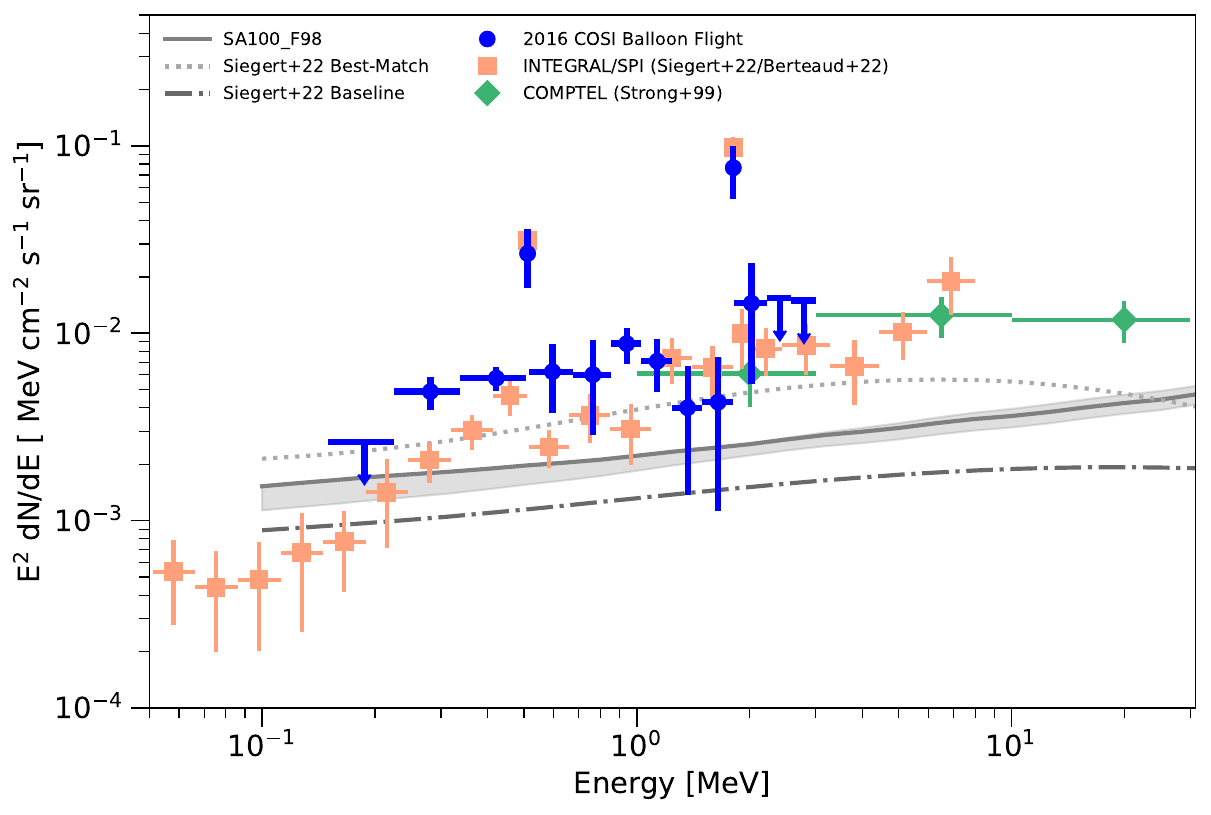}
\caption{Observed intensity towards the inner Galaxy. The blue data points are for COSI, and the error bars are $1\sigma$ statistical only. Upper limits are plotted at the 95\% confidence level. The salmon data points are for INTEGRAL/SPI, and the green data points are for COMPTEL, with the error bars on both measurements including statistical and systematic uncertainty. The grey curves show predictions for our three primary GALPROP models, as specified in the legend. The grey band shows the full range spanned by the six GALPROP models from~\citet{2021arXiv211212745P}, which includes the SA100-F98 model.}
\label{all_spectra}
\end{figure*}

The observed count rate per energy bin from~\citet{Beechert:2022myg} can be converted to flux using
\begin{equation}
\label{eq:flux_main}
    F = \frac{( \frac{N_{\mathrm{obs}}}{\Delta t \times \Delta E})}{\epsilon_{\mathrm{atm}} \times A_{\mathrm{eff}} \times \Delta \Omega},
\end{equation}
or equivalently
\begin{equation}
    F = \bar{F}_{\mathrm{reg}} \times \frac{N_{\mathrm{obs}}}{N_{\mathrm{det}}} \times \epsilon_{\mathrm{atm}} ^{-1}.
\end{equation}
The quantity $N_{\mathrm{obs}}$ is the number of observed photons, and the term $\epsilon_{\mathrm{atm}}$ represents the atmospheric response. 

The atmospheric response can be described in terms of two components. The first component is due to photons that scatter and never reach the detector, thereby causing an attenuation of the original signal. This is already accounted for in the simulations via the transmission probability. However, this does not account for the second component, where photons reach the detector after scattering one or more times. Accounting for this component is especially important when analyzing diffuse sources, as it causes an energy-dependent distortion in the measured spectrum, generally resulting in more photons at lower energy. The exact distortion depends on the spectral index of the source. For an index corresponding to the GDCE ($\sim$1.4), the scatted photons can increase the flux by a factor of $\sim$2 at 0.2 MeV, whereas at 1 MeV there is only a modest increase of $\sim$1.3. We account for this component by applying an energy-dependent correction factor ratio. More details regarding the determination of this correction are provided in Appendix~\ref{sec:appendix}. This ultimately serves as an approximation to the full convolution with the atmospheric response, effectively replacing $\epsilon_\mathrm{atm}$ in Eq.~\ref{eq:flux_main}.

\section{Results}
\label{sec:results}
In Figure~\ref{all_spectra} we show the resulting intensity obtained by converting the observed emission from~\citet{Beechert:2022myg}, as described in the previous sections. The corresponding data is available in Appendix~\ref{sec:data_tables}. For comparison, we also plot similar measurements from INTEGRAL/SPI~\citep{2022A&A...660A.130S,2022PhRvD.106b3030B} and COMPTEL~\citep{1994A&A...292...82S,1999ApL&C..39..209S}. The COMPTEL data has been derived using a maximum-entropy approach\footnote{Details of the derivation and corresponding sky maps for the COMPTEL data can be found at \url{https://www.mpe.mpg.de/~aws/comptel/aws/skymos/skymos.html}.}. This method employs an on-off analysis, where high-latitude observations are used to estimate the background for the inner Galaxy. The INTEGRAL data\footnote{INTEGRAL/SPI data are available at \url{https://zenodo.org/record/7984451}.} shows the total diffuse emission, which is assumed to have contributions from IC emission, the 0.511 MeV line and ortho-positronium, the $^{26}$Al line at 1.809 MeV, and a population of unresolved point sources, mostly cataclysmic variables that contribute below 0.1 MeV (i.e.~these are the templates used to derive the diffuse emission, starting with the raw data). The INTEGRAL and COMPTEL intensities are obtained for the sky region bounded by $|l| \ \mathrm{\&} \ |b| < 47.5^\circ$. Meanwhile, the COSI intensities are obtained for the sky region $|l| < 65^\circ \ \mathrm{\&} \ |b| < 45^\circ$. While the two sky regions are not exactly coincident, they are sufficiently close that the region averaged intensities can be directly compared.

As can be seen in Figure~\ref{all_spectra}, the COSI observations are in overall good agreement with the observations from INTEGRAL/SPI and COMPTEL. Thus, the continuum emission detected in~\citet{Beechert:2022myg} can most likely be attributed to the GDCE, and the results presented in Figure~\ref{all_spectra} provide the first estimate of the observed flux during the 2016 COSI balloon flight.  

\section{Summary, Discussion, and conclusion}
\label{sec:conclusion}
In this work we have determined the flux from the GDCE observed towards the inner Galaxy during the 2016 COSI balloon flight, finding good agreement with previous measurements from INTEGRAL/SPI and COMPTEL. Our measurements are based on the continuum emission initially detected in~\citet{Beechert:2022myg}, which was focused on the detection of $^{26}$Al from the Galaxy. Similar to the analysis of the nuclear line, the high latitude regions chosen for the background model are expected to be relatively clear of IC emission, compared to the bright emission towards the inner Galaxy. Therefore, any related bias in the continuum emission is expected to be minor. However, the background subtraction is complicated by the fact that the balloon altitude continuously changes throughout the flight, and this affects the background normalizaton level. Fully quantifying the systematics associated with the background subtraction is beyond the scope of this paper, but will be the subject of a follow-up work.

The measured COSI spectrum in Figure~\ref{all_spectra} has not been corrected for any contribution from point sources. On the other hand, the INTEGRAL data has removed the resolved point sources using the INTEGRAL/SPI catalog, which contains 256 sources~\citep{2011ApJ...739...29B}. The combined flux from this contribution is most prominent towards COSI's lower energy bound (200 keV), with a flux level of $\sim 1 \times10^{-3} \ \mathrm{MeV \ cm^{-2} \ s^{-1} \ sr^{-1}}$, but it quickly falls off at higher energies. Thus, although subdominant, it may still introduce some bias in the comparison between the COSI and INTEGRAL data sets. We expect that by accounting for this component in the COSI data it will further improve the agreement between the two measurements. 

In addition to resolved sources, the observed flux likely has some contribution coming from unresolved point sources~\citep{2008ApJ...679.1315B,2008ApJ...682..400P,2021ApJ...916...28T,2022arXiv221205713T}. Following a similar approach as~\citet{2021ApJ...916...28T}, we made a simple estimate of this contribution in the COSI band based on matched sources detected by both \textit{Swift}-BAT and \textit{Fermi}-LAT. To estimate the contribution in the COSI energy band we performed a power law extrapolation from the \textit{Swift}-BAT band. The expected contribution is subdominant compared to the GDCE, with a maximum intensity ratio of $I_{\mathrm{unresolved}}/I_{\mathrm{GDCE}} \sim 
 0.3$ at $E = 0.2$ MeV. Nevertheless, investigating these sources in detail will be an important avenue for future studies.  

As revealed by the INTEGRAL measurements, the spectrum of the GDCE has many intriguing features. For one, there is likely a close correlation with the 0.511 MeV line from electron-positron annihilation and corresponding ortho-positronium continuum. There also appears to be a break in the spectrum near $\sim$1 MeV, plausibly indicative of contributions from multiple source classes. Moreover, the observed flux is above the GALPROP prediction by a factor of $\sim 2-3$, as has been discussed since~\citet{2011ApJ...739...29B}. With the current data from the COSI balloon flight we do not expect to have the sensitivity needed to make significant advances in our understanding of the GDCE. But nevertheless, the observations presented in this work are the first diffuse continuum results from COSI, and they serve as a proof of principle. The COSI satellite mission should mark a giant leap forward in this regard. This will be the result of COSI's large field of view and improved sensitivity, as well as advancements in the modeling of the GDCE and data analysis tools, i.e.~cosipy\footnote{\url{https://github.com/cositools/cosipy}}~\citep{2023arXiv230811436M}, currently under development. The COSI satellite mission should enable measurements of a highly precise spectrum over the entire sky, including the inner Galaxy. Such measurements will likely play a major role in disentangling the different components of the emission, including searches for new physics (e.g.~\cite{2022PhRvD.106b3030B,2023JCAP...02..006C}). The measurements in the MeV band will also be intimately connected with other wavelengths, including radio, GeV, and possibly TeV. Finally, by measuring the truly diffuse emission with unprecedented sensitivity in the MeV band, COSI will likely lead to new insights regarding the sources of CR electrons in the Galaxy.     

\section*{Acknowledgements}
The COSI balloon program was supported through NASA APRA grants NNX14AC81G and 80NSSC19K1389. This work is partially supported under NASA contract 80GSFC21C0059, and it is also supported in part by the Centre National d’Etudes Spatiales (CNES). GALPROP development is partially funded via NASA grants 80NSSC22K0477, 80NSSC22K0718, and 80NSSC23K0169. CMK's research was supported by an appointment to the NASA Postdoctoral Program at NASA Goddard Space Flight Center, administered by Oak Ridge Associated Universities under contract with NASA.

\appendix 

\section{Data Tables}
\label{sec:data_tables}
The COSI data presented in this work are provided in Tables~\ref{tab:continuum_data} and~\ref{tab:cosi_intensity_table}.
\begin{deluxetable}{lccc}
\tablecaption{Residual Continuum Emission \label{tab:continuum_data}}
\tablehead{
\colhead{$E$} & \colhead{$\Delta E / 2$}  & \colhead{Rate} & \colhead{1$\sigma$ Rate Error}  \\
\colhead{[MeV]} & \colhead{[MeV]} & \colhead{[$\mathrm{cts \ MeV^{-1} \ s^{-1}}$]} & \colhead{[$\mathrm{cts \ MeV^{-1} \ s^{-1}}$]}
}
\startdata
0.187  & 0.037     & 0.0002  & 0.0009   \\
0.281   & 0.056     & 0.0737  & 0.0146    \\
0.422   & 0.084     & 0.1153  & 0.0172  \\
0.511    & 0.005     & 0.4110 & 0.1425   \\
0.596    & 0.080    & 0.0933  & 0.0374    \\
0.764   & 0.089     & 0.0691 & 0.0361    \\
0.940    & 0.088    & 0.0675  & 0.0147    \\
1.133   & 0.106     & 0.0367  & 0.0114    \\
1.367   & 0.128     & 0.0128  & 0.0085   \\
1.648    & 0.154     & 0.0080 & 0.0059   \\
1.809    & 0.007    & 0.0881  & 0.0277    \\
2.023   & 0.207     & 0.0121 & 0.0076    \\
2.408    & 0.178    & 0.0005  & 0.0019    \\
2.793   & 0.206     & 0.0004  & 0.0011    \\
\enddata
\tablecomments{Data from~\citet{Beechert:2022myg}, as plotted in Figure~\ref{fig:sims}.}
\end{deluxetable}

\begin{deluxetable}{lccc}
\tablecaption{Measured Intensity \label{tab:cosi_intensity_table}}
\tablehead{
\colhead{$E$} & \colhead{$\Delta E / 2$}  & \colhead{Intensity} & \colhead{1$\sigma$ Intensity Error}  \\
\colhead{[MeV]} & \colhead{[MeV]} & \colhead{[$\mathrm{MeV \ cm^{-2} \ s^{-1} \ sr^{-1}}$]} & \colhead{[$\mathrm{MeV \ cm^{-2} \ s^{-1} \ sr^{-1}}$]}
}
\startdata
0.187  & 0.037     & 0.0003  & 0.0012   \\
0.281   & 0.056     & 0.0049  & 0.0010    \\
0.422   & 0.084     & 0.0058  & 0.0009  \\
0.511    & 0.005     & 0.0267 & 0.0092   \\
0.596    & 0.080    & 0.0062  & 0.0025    \\
0.764   & 0.089     & 0.0060 & 0.0031    \\
0.940    & 0.088    & 0.0088  & 0.0019    \\
1.133   & 0.106     & 0.0071  & 0.0022    \\
1.367   & 0.128     & 0.0040  & 0.0026   \\
1.648    & 0.154     & 0.0043 & 0.0032   \\
1.809    & 0.007    & 0.0765  & 0.0241    \\
2.023   & 0.207     & 0.0145 & 0.0091    \\
2.408    & 0.178    & 0.0017  & 0.0069    \\
2.793   & 0.206     & 0.0029  & 0.0089    \\
\enddata
\tablecomments{Measured intensity from the 2016 COSI balloon flight, as plotted in Figure~\ref{all_spectra}.}
\end{deluxetable}

\section{Atmospheric Response}
\label{sec:appendix}

As $\gamma$ rays with energies between $\sim$ 0.1 - 10 MeV travel through Earth's atmosphere they may undergo Compton scattering, which causes attenuation and distortion of the original signal. The scattering of MeV photons in the atmosphere can be divided into two components. First, a fraction of photons from a source will travel through the atmosphere and reach the detector without scattering, which we refer to as the transmitted component. This component can be accounted for in a fairly straightforward manner by convolving the detector response with the transmission probability (TP). Generally, the TP depends on the initial energy of the photon, the off-axis angle of the source, and the altitude of the observations. Indeed, such a method is employed in our MEGAlib simulations. However, this does not account for the second component, where photons from a source will reach the detector after one or more scatters, which we refer to as the scattered component. Accounting for this component is much more challenging, as it requires tracking the $\gamma$-ray transport in the atmosphere for all incident photons.

The scattered component is important to account for when analyzing balloon-borne observations, as it can produce a significant energy-dependent distortion in the measured spectrum. In general, this tends to lead to more photons towards lower energies. For diffuse continuum sources, such as the GDCE, photons enter the detector from all directions, and therefore the spectral distortion from the scattered photons becomes very significant. The effect is not as crucial for point sources observed with imaging telescopes, such as COSI, because they have the ability to decipher the photon's incident direction. However, this also has a strong dependence on the instrument's angular resolution. 

A full analysis of the response for $\gamma$-ray transport in the atmosphere was recently made by~\citet{karwin_atm} (in prep). We use results from that work to correct the spectrum of the GDCE for the scattered component. Here we briefly summarize the main points of the calculation. The atmospheric response is simulated with the COSI atmosphere pipeline\footnote{\url{https://github.com/cositools/cosi-atmosphere}}, which employs MEGAlib. An atmospheric model is created using the latest version (v2.1) of the Naval Research Laboratory's Mass Spectrometer Incoherent Scatter Radar Model (NRLMSIS)\footnote{\url{https://swx-trec.com/msis}}~\citep{2002JGRA..107.1468P,2021E&SS....801321E,2022JGRA..12730896E}, implemented in the COSI atmosphere pipeline via the python interface, \textit{pymsis}\footnote{\url{https://swxtrec.github.io/pymsis/}}. The atmospheric model specifies the altitude profile of the number density for the primary species of the atmosphere (i.e.~nitrogen, oxygen, argon, and helium). For the model we use a representative date, geographical location, and altitude from the COSI balloon flight, namely, 2016-06-13, $(\mathrm{lat,lon})=(-5.66^\circ,-107.38^\circ)$, and $33.5$ km, respectively. The atmospheric profile is used to construct a spherical mass model for the simulations, extending from Earth's surface\footnote{We use Earth's equatorial radius, which is slightly larger than the polar radius of 6357 km.} ($R_\Earth=6378$ km) to an altitude of 200 km. An isotropic source is simulated with a flat energy spectrum (i.e. constant number of photons per energy bin) between 0.01 - 10 MeV using $10^7$ triggers. With the isotropic source, photons are directed towards Earth's surface at all incident angles. The simulations track the $\gamma$-ray transport through the atmosphere, including initial and final values of the photon's energy, position, and direction. The final values are tracked for a watched altitude of 33.5 km. This information is then used to construct the atmospheric response matrices. 

\begin{figure}[t]
\centering
\includegraphics[width=0.6\columnwidth]{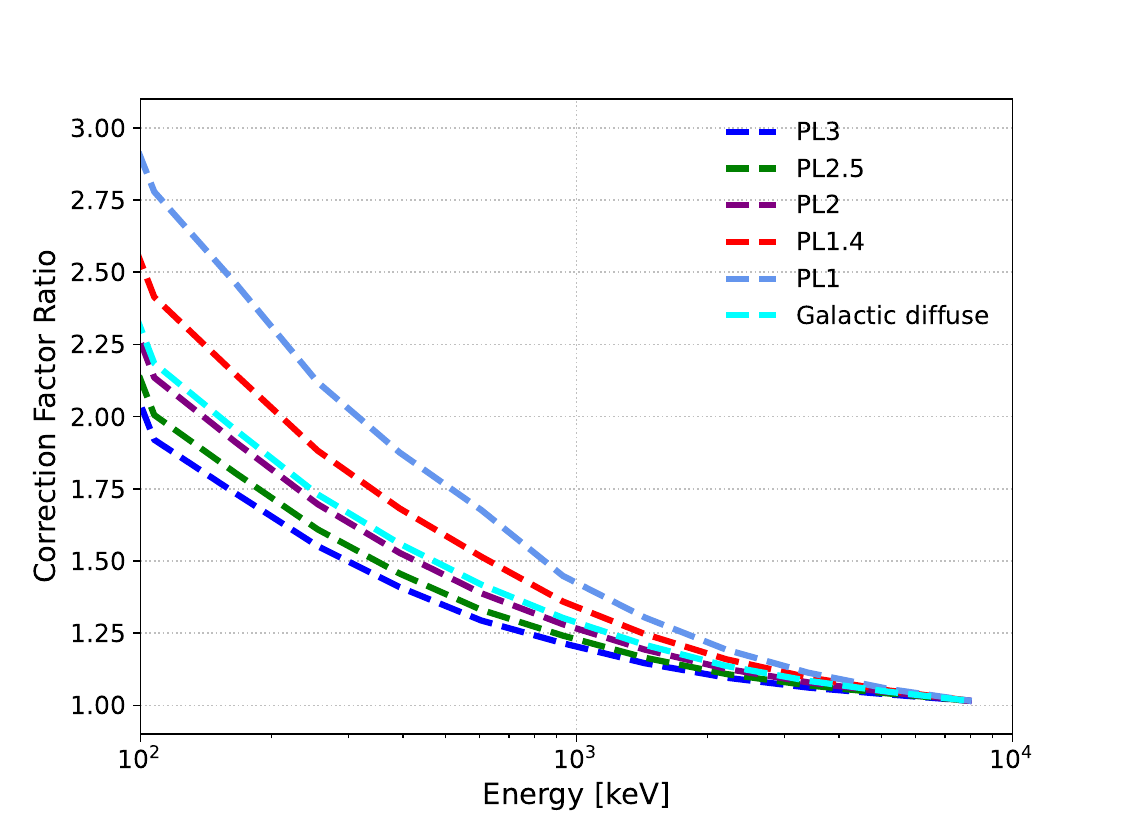}
\caption{Atmospheric correction factor ratio for an off-axis angle of $44^\circ$. We consider a power law (PL) spectral model with different spectral indices, as specified in the legend, as well as the model of the Galactic diffuse (SA100-F98) from the simulations.}
\label{fig:correction}
\end{figure}

In calculating COSI's response to the GDCE, we have already accounted for the transmission probability of the transmitted component; therefore, we only need to correct for the scattered component, as discussed above. To do this we multiply the effective area by an energy-dependent correction factor ratio, $R$, defined as 
\begin{equation}
    R = \frac{c_\mathrm{tot}}{c_\mathrm{tran}},
\end{equation}
where $c_\mathrm{tran}$ is the correction factor for the transmitted photons only (equivalent to the transmission probability) and $c_\mathrm{tot}$ is the total correction factor that includes both the transmitted and scattered components. The correction factors are defined as the ratio of the predicted counts to the model counts, where the latter are obtained by forward-folding the model with the atmospheric response. The correction factor ratio serves as an approximation to the full convolution with $\epsilon_{\mathrm{atm}}$, as specified in Eq.~\ref{eq:flux_main} (i.e.~$R$ replaces $\epsilon_{\mathrm{atm}}$). For the calculation we use an off-axis angle of $44.1^\circ$, which is the mean within the signal region, with respect to the instrument pointing during the observations. For the forward-folding we use a power law spectral model with a photon index of 1.4. This is the best-fit value ($1.4 \pm 0.19$) from~\citet{2022A&A...660A.130S}, based on the INTEGRAL/SPI measurements in the energy range 0.5 - 8 MeV. The corresponding correction factor ratio is shown in Figure~\ref{fig:correction}, where we also show how the ratio changes for different assumptions of the spectral index. In general, the correction factor is larger for harder sources, as there are more photons at higher energies that get scattered down to the lower energy band. For comparison, we also show the Galactic diffuse model that was used for the simulations. At 200 keV the difference in the correction factor ratio with respect to a spectral index of 1.4 is $\sim$10\%. An additional $\sim 10 - 20 \%$ uncertainty arises from our assumption on the viewing angle. Thus, these assumptions have a minor impact on our results.


\bibliography{sample631}

\begin{thebibliography}{}
\expandafter\ifx\csname natexlab\endcsname\relax\def\natexlab#1{#1}\fi
\providecommand{\url}[1]{\href{#1}{#1}}
\providecommand{\dodoi}[1]{doi:~\href{http://doi.org/#1}{\nolinkurl{#1}}}
\providecommand{\doeprint}[1]{\href{http://ascl.net/#1}{\nolinkurl{http://ascl.net/#1}}}
\providecommand{\doarXiv}[1]{\href{https://arxiv.org/abs/#1}{\nolinkurl{https://arxiv.org/abs/#1}}}

\bibitem[{Agostinelli {et~al.}(2003)Agostinelli, Allison, Amako, Apostolakis,
  Araujo, Arce, Asai, Axen, Banerjee, Barrand, Behner, Bellagamba, Boudreau,
  Broglia, Brunengo, Burkhardt, Chauvie, Chuma, Chytracek, Cooperman, Cosmo,
  Degtyarenko, Dell'Acqua, Depaola, Dietrich, Enami, Feliciello, Ferguson,
  Fesefeldt, Folger, Foppiano, Forti, Garelli, Giani, Giannitrapani, Gibin,
  {Gómez Cadenas}, González, {Gracia Abril}, Greeniaus, Greiner, Grichine,
  Grossheim, Guatelli, Gumplinger, Hamatsu, Hashimoto, Hasui, Heikkinen,
  Howard, Ivanchenko, Johnson, Jones, Kallenbach, Kanaya, Kawabata, Kawabata,
  Kawaguti, Kelner, Kent, Kimura, Kodama, Kokoulin, Kossov, Kurashige, Lamanna,
  Lampén, Lara, Lefebure, Lei, Liendl, Lockman, Longo, Magni, Maire,
  Medernach, Minamimoto, {Mora de Freitas}, Morita, Murakami, Nagamatu,
  Nartallo, Nieminen, Nishimura, Ohtsubo, Okamura, O'Neale, Oohata, Paech,
  Perl, Pfeiffer, Pia, Ranjard, Rybin, Sadilov, {Di Salvo}, Santin, Sasaki,
  Savvas, Sawada, Scherer, Sei, Sirotenko, Smith, Starkov, Stoecker, Sulkimo,
  Takahata, Tanaka, Tcherniaev, {Safai Tehrani}, Tropeano, Truscott, Uno,
  Urban, Urban, Verderi, Walkden, Wander, Weber, Wellisch, Wenaus, Williams,
  Wright, Yamada, Yoshida, \& Zschiesche}]{geant4_2003}
Agostinelli, S., Allison, J., Amako, K., {et~al.} 2003, Nuclear Instruments and
  Methods in Physics Research Section A: Accelerators, Spectrometers, Detectors
  and Associated Equipment, 506, 250,
  \dodoi{https://doi.org/10.1016/S0168-9002(03)01368-8}

\bibitem[{{Ajello} {et~al.}(2012){Ajello}, {Shaw}, {Romani}, {Dermer},
  {Costamante}, {King}, {Max-Moerbeck}, {Readhead}, {Reimer}, {Richards}, \&
  {Stevenson}}]{2012ApJ...751..108A}
{Ajello}, M., {Shaw}, M.~S., {Romani}, R.~W., {et~al.} 2012, \apj, 751, 108,
  \dodoi{10.1088/0004-637X/751/2/108}

\bibitem[{Beechert {et~al.}(2022)}]{Beechert:2022myg}
Beechert, J., {et~al.} 2022, \apj, 928, 119, \dodoi{10.3847/1538-4357/ac56dc}

\bibitem[{{Beechert} {et~al.}(2022){Beechert}, {Lazar}, {Boggs}, {Brandt},
  {Chang}, {Chu}, {Gulick}, {Kierans}, {Lowell}, {Pellegrini}, {Roberts},
  {Siegert}, {Sleator}, {Tomsick}, \& {Zoglauer}}]{BeechertCOSICalib}
{Beechert}, J., {Lazar}, H., {Boggs}, S.~E., {et~al.} 2022, Nuclear Instruments
  and Methods in Physics Research A, 1031, 166510,
  \dodoi{10.1016/j.nima.2022.166510}

\bibitem[{{Berteaud} {et~al.}(2022){Berteaud}, {Calore}, {Iguaz}, {Serpico}, \&
  {Siegert}}]{2022PhRvD.106b3030B}
{Berteaud}, J., {Calore}, F., {Iguaz}, J., {Serpico}, P.~D., \& {Siegert}, T.
  2022, \prd, 106, 023030, \dodoi{10.1103/PhysRevD.106.023030}

\bibitem[{{Bisschoff} {et~al.}(2019){Bisschoff}, {Potgieter}, \&
  {Aslam}}]{2019ApJ...878...59B}
{Bisschoff}, D., {Potgieter}, M.~S., \& {Aslam}, O.~P.~M. 2019, \apj, 878, 59,
  \dodoi{10.3847/1538-4357/ab1e4a}

\bibitem[{{Boggs} \& {Jean}(2000)}]{2000A&AS..145..311B}
{Boggs}, S.~E., \& {Jean}, P. 2000, \aaps, 145, 311,
  \dodoi{10.1051/aas:2000107}

\bibitem[{{Boggs} {et~al.}(2000){Boggs}, {Lin}, {Slassi-Sennou}, {Coburn}, \&
  {Pelling}}]{2000ApJ...544..320B}
{Boggs}, S.~E., {Lin}, R.~P., {Slassi-Sennou}, S., {Coburn}, W., \& {Pelling},
  R.~M. 2000, \apj, 544, 320, \dodoi{10.1086/317167}

\bibitem[{{Bouchet} {et~al.}(2008){Bouchet}, {Jourdain}, {Roques}, {Strong},
  {Diehl}, {Lebrun}, \& {Terrier}}]{2008ApJ...679.1315B}
{Bouchet}, L., {Jourdain}, E., {Roques}, J.~P., {et~al.} 2008, \apj, 679, 1315,
  \dodoi{10.1086/529489}

\bibitem[{{Bouchet} {et~al.}(2011){Bouchet}, {Strong}, {Porter}, {Moskalenko},
  {Jourdain}, \& {Roques}}]{2011ApJ...739...29B}
{Bouchet}, L., {Strong}, A.~W., {Porter}, T.~A., {et~al.} 2011, \apj, 739, 29,
  \dodoi{10.1088/0004-637X/739/1/29}

\bibitem[{{Caputo} {et~al.}(2023){Caputo}, {Negro}, {Regis}, \&
  {Taoso}}]{2023JCAP...02..006C}
{Caputo}, A., {Negro}, M., {Regis}, M., \& {Taoso}, M. 2023, \jcap, 2023, 006,
  \dodoi{10.1088/1475-7516/2023/02/006}

\bibitem[{Caputo {et~al.}(2022)}]{Caputo:2022xpx}
Caputo, R., {et~al.} 2022, J. Astron. Telesc. Instrum. Syst., 8, 044003,
  \dodoi{10.1117/1.JATIS.8.4.044003}

\bibitem[{{Cumani} {et~al.}(2019){Cumani}, {Hernanz}, {Kiener}, {Tatischeff},
  \& {Zoglauer}}]{2019MmSAI..90..226C}
{Cumani}, P., {Hernanz}, M., {Kiener}, J., {Tatischeff}, V., \& {Zoglauer}, A.
  2019, \memsai, 90, 226

\bibitem[{{Cummings} {et~al.}(2016){Cummings}, {Stone}, {Heikkila}, {Lal},
  {Webber}, {J{\'o}hannesson}, {Moskalenko}, {Orlando}, \&
  {Porter}}]{2016ApJ...831...18C}
{Cummings}, A.~C., {Stone}, E.~C., {Heikkila}, B.~C., {et~al.} 2016, \apj, 831,
  18, \dodoi{10.3847/0004-637X/831/1/18}

\bibitem[{{Emmert} {et~al.}(2021){Emmert}, {Drob}, {Picone}, {Siskind},
  {Jones}, {Mlynczak}, {Bernath}, {Chu}, {Doornbos}, {Funke}, {Goncharenko},
  {Hervig}, {Schwartz}, {Sheese}, {Vargas}, {Williams}, \&
  {Yuan}}]{2021E&SS....801321E}
{Emmert}, J.~T., {Drob}, D.~P., {Picone}, J.~M., {et~al.} 2021, Earth and Space
  Science, 8, e01321, \dodoi{10.1029/2020EA001321}

\bibitem[{{Emmert} {et~al.}(2022){Emmert}, {Jones}, {Siskind}, {Drob},
  {Picone}, {Stevens}, {Bailey}, {Bender}, {Bernath}, {Funke}, {Hervig}, \&
  {P{\'e}rot}}]{2022JGRA..12730896E}
{Emmert}, J.~T., {Jones}, M., {Siskind}, D.~E., {et~al.} 2022, Journal of
  Geophysical Research (Space Physics), 127, e2022JA030896,
  \dodoi{10.1029/2022JA030896}

\bibitem[{{Gehrels} \& {Tueller}(1993)}]{1993ApJ...407..597G}
{Gehrels}, N., \& {Tueller}, J. 1993, \apj, 407, 597, \dodoi{10.1086/172541}

\bibitem[{{Harris} {et~al.}(1990){Harris}, {Share}, {Leising}, {Kinzer}, \&
  {Messina}}]{1990ApJ...362..135H}
{Harris}, M.~J., {Share}, G.~H., {Leising}, M.~D., {Kinzer}, R.~L., \&
  {Messina}, D.~C. 1990, \apj, 362, 135, \dodoi{10.1086/169250}

\bibitem[{Johannesson {et~al.}(2018)Johannesson, Porter, \&
  Moskalenko}]{Johannesson:2018bit}
Johannesson, G., Porter, T.~A., \& Moskalenko, I.~V. 2018, Astrophys. J., 856,
  45, \dodoi{10.3847/1538-4357/aab26e}

\bibitem[{{Johnson} \& {Haymes}(1973)}]{1973ApJ...184..103J}
{Johnson}, W.~N., I., \& {Haymes}, R.~C. 1973, \apj, 184, 103,
  \dodoi{10.1086/152309}

\bibitem[{{Karwin} {et~al.}(2023){Karwin}, {Kierans}, {Shih}, {Romani},
  {Dermer}, \& {Costamante}}]{karwin_atm}
{Karwin}, C., {Kierans}, C., {Shih}, A., {et~al.} 2023, in prep

\bibitem[{Kierans {et~al.}(2016)}]{Kierans:2016qik}
Kierans, C.~A., {et~al.} 2016, PoS, INTEGRAL2016, 075,
  \dodoi{10.22323/1.285.0075}

\bibitem[{Kierans {et~al.}(2020)}]{Kierans:2019aqz}
---. 2020, \apj, 895, 44, \dodoi{10.3847/1538-4357/ab89a9}

\bibitem[{{Kinzer} {et~al.}(1999){Kinzer}, {Purcell}, \&
  {Kurfess}}]{1999ApJ...515..215K}
{Kinzer}, R.~L., {Purcell}, W.~R., \& {Kurfess}, J.~D. 1999, \apj, 515, 215,
  \dodoi{10.1086/306997}

\bibitem[{{Krivonos} {et~al.}(2007){Krivonos}, {Revnivtsev}, {Churazov},
  {Sazonov}, {Grebenev}, \& {Sunyaev}}]{2007A&A...463..957K}
{Krivonos}, R., {Revnivtsev}, M., {Churazov}, E., {et~al.} 2007, \aap, 463,
  957, \dodoi{10.1051/0004-6361:20065626}

\bibitem[{{Leventhal} {et~al.}(1978){Leventhal}, {MacCallum}, \&
  {Stang}}]{1978ApJ...225L..11L}
{Leventhal}, M., {MacCallum}, C.~J., \& {Stang}, P.~D. 1978, \apjl, 225, L11,
  \dodoi{10.1086/182782}

\bibitem[{{Lowell} {et~al.}(2017){Lowell}, {Boggs}, {Chiu}, {Kierans},
  {Sleator}, {Tomsick}, {Zoglauer}, {Chang}, {Tseng}, {Yang}, {Jean}, {von
  Ballmoos}, {Lin}, \& {Amman}}]{2017ApJ...848..119L}
{Lowell}, A.~W., {Boggs}, S.~E., {Chiu}, C.~L., {et~al.} 2017, \apj, 848, 119,
  \dodoi{10.3847/1538-4357/aa8ccb}

\bibitem[{Mahoney {et~al.}(1984)Mahoney, Ling, Wheaton, \&
  Jacobson}]{mahoney1984heao}
Mahoney, W., Ling, J., Wheaton, W.~A., \& Jacobson, A. 1984, \apj, 286, 578

\bibitem[{{Martinez-Castellanos} {et~al.}(2023){Martinez-Castellanos},
  {Gallego}, {Huang}, {Karwin}, {Kierans}, {Lommler}, {Mittal}, {Negro},
  {Neights}, {Pike}, {Sheng}, {Siegert}, {Yoneda}, {Zoglauer}, {Tomsick},
  {Boggs}, {Hartmann}, {Ajello}, {Burns}, {Fryer}, {Lowell}, {Malzac},
  {Roberts}, {Saint-Hilaire}, {Shih}, {Sleator}, {Takahashi}, {Tavecchio},
  {Wulf}, {Beechert}, {Gulick}, {Joens}, {Lazar}, {Martinez Oliveros},
  {Matsumoto}, {Melia}, {Amman}, {Bal}, {von Ballmoos}, {Bates},
  {B{\"o}ttcher}, {Bulgarelli}, {Cavazzuti}, {Chang}, {Chen}, {Chu},
  {Ciabattoni}, {Costamante}, {Dreyer}, {Fioretti}, {Fenu}, {Ghirlanda},
  {Grove}, {Jean}, {Khatiya}, {Kn{\"o}dlseder}, {Krause}, {Leising}, {Lewis},
  {Marcotulli}, {Al Nussirat}, {Nakazawa}, {Oberlack}, {Palmore}, {Panebianco},
  {Parmiggiani}, {Parsotan}, {Rogers}, {Schutte}, {Smale}, {Smith}, {Trigg},
  {Venters}, {Watanabe}, \& {Zhang}}]{2023arXiv230811436M}
{Martinez-Castellanos}, I., {Gallego}, S., {Huang}, C.-Y., {et~al.} 2023, arXiv
  e-prints, arXiv:2308.11436, \dodoi{10.48550/arXiv.2308.11436}

\bibitem[{{Orlando}(2018)}]{2018MNRAS.475.2724O}
{Orlando}, E. 2018, \mnras, 475, 2724, \dodoi{10.1093/mnras/stx3280}

\bibitem[{{Picone} {et~al.}(2002){Picone}, {Hedin}, {Drob}, \&
  {Aikin}}]{2002JGRA..107.1468P}
{Picone}, J.~M., {Hedin}, A.~E., {Drob}, D.~P., \& {Aikin}, A.~C. 2002, Journal
  of Geophysical Research (Space Physics), 107, 1468,
  \dodoi{10.1029/2002JA009430}

\bibitem[{Porter {et~al.}(2017)Porter, Johannesson, \&
  Moskalenko}]{porter2017high}
Porter, T.~A., Johannesson, G., \& Moskalenko, I.~V. 2017, Astrophys. J., 846,
  23pp

\bibitem[{{Porter} {et~al.}(2021){Porter}, {Johannesson}, \&
  {Moskalenko}}]{2021arXiv211212745P}
{Porter}, T.~A., {Johannesson}, G., \& {Moskalenko}, I.~V. 2021, arXiv
  e-prints, arXiv:2112.12745, \dodoi{10.48550/arXiv.2112.12745}

\bibitem[{{Porter} {et~al.}(2022){Porter}, {J{\'o}hannesson}, \&
  {Moskalenko}}]{2022ApJS..262...30P}
{Porter}, T.~A., {J{\'o}hannesson}, G., \& {Moskalenko}, I.~V. 2022, \apjs,
  262, 30, \dodoi{10.3847/1538-4365/ac80f6}

\bibitem[{{Porter} {et~al.}(2008){Porter}, {Moskalenko}, {Strong}, {Orlando},
  \& {Bouchet}}]{2008ApJ...682..400P}
{Porter}, T.~A., {Moskalenko}, I.~V., {Strong}, A.~W., {Orlando}, E., \&
  {Bouchet}, L. 2008, \apj, 682, 400, \dodoi{10.1086/589615}

\bibitem[{{Porter} \& {Strong}(2005)}]{2005ICRC....4...77P}
{Porter}, T.~A., \& {Strong}, A.~W. 2005, in International Cosmic Ray
  Conference, Vol.~4, 29th International Cosmic Ray Conference (ICRC29), Volume
  4, 77, \dodoi{10.48550/arXiv.astro-ph/0507119}

\bibitem[{{Schoenfelder} {et~al.}(1993){Schoenfelder}, {Aarts}, {Bennett}, {de
  Boer}, {Clear}, {Collmar}, {Connors}, {Deerenberg}, {Diehl}, {von Dordrecht},
  {den Herder}, {Hermsen}, {Kippen}, {Kuiper}, {Lichti}, {Lockwood}, {Macri},
  {McConnell}, {Morris}, {Much}, {Ryan}, {Simpson}, {Snelling}, {Stacy},
  {Steinle}, {Strong}, {Swanenburg}, {Taylor}, {de Vries}, \&
  {Winkler}}]{1993ApJS...86..657S}
{Schoenfelder}, V., {Aarts}, H., {Bennett}, K., {et~al.} 1993, \apjs, 86, 657,
  \dodoi{10.1086/191794}

\bibitem[{{Siegert} {et~al.}(2022){Siegert}, {Berteaud}, {Calore}, {Serpico},
  \& {Weinberger}}]{2022A&A...660A.130S}
{Siegert}, T., {Berteaud}, J., {Calore}, F., {Serpico}, P.~D., \& {Weinberger},
  C. 2022, \aap, 660, A130, \dodoi{10.1051/0004-6361/202142639}

\bibitem[{Siegert {et~al.}(2020)}]{Siegert:2020oxw}
Siegert, T., {et~al.} 2020, \apj, 897, 45, \dodoi{10.3847/1538-4357/ab9607}

\bibitem[{{Skibo} {et~al.}(1995){Skibo}, {Ramaty}, \&
  {Purcell}}]{1995ICRC....2..219S}
{Skibo}, J.~G., {Ramaty}, R., \& {Purcell}, W.~R. 1995, in International Cosmic
  Ray Conference, Vol.~2, International Cosmic Ray Conference, 219

\bibitem[{{Skibo} {et~al.}(1997){Skibo}, {Johnson}, {Kurfess}, {Kinzer},
  {Jung}, {Grove}, {Purcell}, {Ulmer}, {Gehrels}, \&
  {Tueller}}]{1997ApJ...483L..95S}
{Skibo}, J.~G., {Johnson}, W.~N., {Kurfess}, J.~D., {et~al.} 1997, \apjl, 483,
  L95, \dodoi{10.1086/310755}

\bibitem[{{Stecker}(1971)}]{1971NASSP.249.....S}
{Stecker}, F.~W. 1971, {Cosmic gamma rays}, Vol. 249

\bibitem[{Strong \& Collmar(2019)}]{Strong:2019yfj}
Strong, A., \& Collmar, W. 2019, Mem. Soc. Ast. It., 90, 297.
\newblock \doarXiv{1907.07454}

\bibitem[{{Strong}(2011)}]{2011crpa.conf..473S}
{Strong}, A.~W. 2011, in Cosmic Rays for Particle and Astroparticle Physics,
  ed. S.~{Giani}, C.~{Leroy}, \& P.~G. {Rancoita}, 473--481,
  \dodoi{10.1142/9789814329033_0059}

\bibitem[{{Strong} {et~al.}(1999){Strong}, {Bloemen}, {Diehl}, {Hermsen}, \&
  {Sch{\"o}nfelder}}]{1999ApL&C..39..209S}
{Strong}, A.~W., {Bloemen}, H., {Diehl}, R., {Hermsen}, W., \&
  {Sch{\"o}nfelder}, V. 1999, Astrophysical Letters and Communications, 39,
  209, \dodoi{10.48550/arXiv.astro-ph/9811211}

\bibitem[{{Strong} {et~al.}(1994){Strong}, {Bennett}, {Bloemen}, {Diehl},
  {Hermsen}, {Morris}, {Schoenfelder}, {Stacy}, {de Vries}, {Varendorff},
  {Winkler}, \& {Youssefi}}]{1994A&A...292...82S}
{Strong}, A.~W., {Bennett}, K., {Bloemen}, H., {et~al.} 1994, \aap, 292, 82

\bibitem[{{Strong} {et~al.}(1996){Strong}, {Bennett}, {Bloemen}, {Diehl},
  {Hermsen}, {Purcell}, {Schoenfelder}, {Stacy}, {Winkler}, \&
  {Youssefi}}]{1996A&AS..120C.381S}
---. 1996, \aaps, 120, 381

\bibitem[{{Teegarden} {et~al.}(1985){Teegarden}, {Cline}, {Gehrels}, {Porreca},
  {Tueller}, {Leventhal}, {Huters}, {MacCallum}, \&
  {Stang}}]{1985ICRC....3..307T}
{Teegarden}, B.~J., {Cline}, T.~L., {Gehrels}, N., {et~al.} 1985, in
  International Cosmic Ray Conference, Vol.~3, 19th International Cosmic Ray
  Conference (ICRC19), Volume 3, 307--310

\bibitem[{{Tomsick} {et~al.}(2019){Tomsick}, {Zoglauer}, {Sleator}, {Lazar},
  {Beechert}, {Boggs}, {Roberts}, {Siegert}, {Lowell}, {Wulf}, {Grove},
  {Phlips}, {Brandt}, {Smale}, {Kierans}, {Burns}, {Hartmann}, {Leising},
  {Ajello}, {Fryer}, {Amman}, {Chang}, {Jean}, \& {von
  Ballmoos}}]{COSIofficial}
{Tomsick}, J., {Zoglauer}, A., {Sleator}, C., {et~al.} 2019, in Bulletin of the
  American Astronomical Society, Vol.~51, 98.
\newblock \doarXiv{1908.04334}

\bibitem[{{Tomsick} {et~al.}(2023){Tomsick}, {Boggs}, {Zoglauer}, {Hartmann},
  {Ajello}, {Burns}, {Fryer}, {Karwin}, {Kierans}, {Lowell}, {Malzac},
  {Roberts}, {Saint-Hilaire}, {Shih}, {Siegert}, {Sleator}, {Takahashi},
  {Tavecchio}, {Wulf}, {Beechert}, {Gulick}, {Joens}, {Lazar}, {Neights},
  {Martinez Oliveros}, {Matsumoto}, {Melia}, {Yoneda}, {Amman}, {Bal}, {von
  Ballmoos}, {Bates}, {B{\"o}ttcher}, {Bulgarelli}, {Cavazzuti}, {Chang},
  {Chen}, {Chu}, {Ciabattoni}, {Costamante}, {Dreyer}, {Fioretti}, {Fenu},
  {Gallego}, {Ghirlanda}, {Grove}, {Huang}, {Jean}, {Khatiya},
  {Kn{\"o}dlseder}, {Krause}, {Leising}, {Lewis}, {Lommler}, {Marcotulli},
  {Martinez-Castellanos}, {Mittal}, {Negro}, {Al Nussirat}, {Nakazawa},
  {Oberlack}, {Palmore}, {Panebianco}, {Parmiggiani}, {Parsotan}, {Pike},
  {Rogers}, {Schutte}, {Sheng}, {Smale}, {Smith}, {Trigg}, {Venters},
  {Watanabe}, \& {Zhang}}]{2023arXiv230812362T}
{Tomsick}, J.~A., {Boggs}, S.~E., {Zoglauer}, A., {et~al.} 2023, arXiv
  e-prints, arXiv:2308.12362, \dodoi{10.48550/arXiv.2308.12362}

\bibitem[{{Tsuji} {et~al.}(2022){Tsuji}, {Inoue}, {Yoneda}, {Mukherjee}, \&
  {Odaka}}]{2022arXiv221205713T}
{Tsuji}, N., {Inoue}, Y., {Yoneda}, H., {Mukherjee}, R., \& {Odaka}, H. 2022,
  arXiv e-prints, arXiv:2212.05713, \dodoi{10.48550/arXiv.2212.05713}

\bibitem[{{Tsuji} {et~al.}(2021){Tsuji}, {Yoneda}, {Inoue}, {Aramaki},
  {Karagiorgi}, {Mukherjee}, \& {Odaka}}]{2021ApJ...916...28T}
{Tsuji}, N., {Yoneda}, H., {Inoue}, Y., {et~al.} 2021, \apj, 916, 28,
  \dodoi{10.3847/1538-4357/ac0341}

\bibitem[{{Wang} {et~al.}(2007){Wang}, {Harris}, {Diehl}, {Halloin}, {Cordier},
  {Strong}, {Kretschmer}, {Kn{\"o}dlseder}, {Jean}, {Lichti}, {Roques},
  {Schanne}, {von Kienlin}, {Weidenspointner}, \&
  {Wunderer}}]{2007A&A...469.1005W}
{Wang}, W., {Harris}, M.~J., {Diehl}, R., {et~al.} 2007, \aap, 469, 1005,
  \dodoi{10.1051/0004-6361:20066982}

\bibitem[{{Wang} {et~al.}(2020){Wang}, {Siegert}, {Dai}, {Diehl}, {Greiner},
  {Heger}, {Krause}, {Lang}, {Pleintinger}, \& {Zhang}}]{2020ApJ...889..169W}
{Wang}, W., {Siegert}, T., {Dai}, Z.~G., {et~al.} 2020, \apj, 889, 169,
  \dodoi{10.3847/1538-4357/ab6336}

\bibitem[{{Zoglauer} {et~al.}(2006){Zoglauer}, {Andritschke}, \&
  {Schopper}}]{2006NewAR..50..629Z}
{Zoglauer}, A., {Andritschke}, R., \& {Schopper}, F. 2006, NAR, 50, 629,
  \dodoi{10.1016/j.newar.2006.06.049}

\bibitem[{Zoglauer {et~al.}(2021)}]{Zoglauer:2021coa}
Zoglauer, A., {et~al.} 2021.
\newblock \doarXiv{2102.13158}

\bibitem[{{Zoglauer}(2006)}]{2006PhDT.........3Z}
{Zoglauer}, A.~C. 2006, PhD thesis, -

\end{thebibliography}
\bibliographystyle{aasjournal}



\end{document}